\batchmode
\makeatletter
\def\input@path{{"/home/jacob/Documents/Work/My Papers/2025-Koopman-von Neumann Misattributed/"}}
\makeatother
\documentclass[11pt,twoside,english]{article}
\usepackage[T1]{fontenc}
\usepackage{textcomp}
\usepackage[latin9]{inputenc}
\usepackage{color}
\definecolor{lyxboxbgcolor}{rgb}{0.980469, 0.941406, 0.902344}
\usepackage{babel}
\usepackage{amsmath}
\usepackage{amssymb}
\usepackage{geometry}
\usepackage{setspace}
\usepackage[pdftex,
 bookmarks=true,bookmarksnumbered=true,bookmarksopen=false,
 breaklinks=false,pdfborder={0 0 0},pdfborderstyle={},backref=false,colorlinks=false]
 {hyperref}
\hypersetup{
 pdfauthor={Jacob A. Barandes},
 pdfsubject={Physics},
 pdfkeywords={physics}}

\makeatletter

\let\originalleft\left
\let\originalright\right
\renewcommand{\left}{\mathopen{}\mathclose\bgroup\originalleft}
\renewcommand{\right}{\aftergroup\egroup\originalright}

\def\smalloverbrace#1{\mathop{\vbox{\m@th\ialign{##\crcr%
      \noalign{\kern3\p@}%
      \tiny\downbracefill\crcr\noalign{\kern3\p@\nointerlineskip}%
      $\hfil\displaystyle{#1}\hfil$\crcr}}}\limits}

\def\smallunderbrace#1{\mathop{\vtop{\m@th\ialign{##\crcr
   $\hfil\displaystyle{#1}\hfil$\crcr
   \noalign{\kern3\p@\nointerlineskip}%
   \tiny\upbracefill\crcr\noalign{\kern3\p@}}}}\limits}

\DeclareMathAlphabet{\mymathbb}{U}{bbold}{m}{n}

\makeatother

\begin{document}
\title{The History of Hilbert-Space Formulations of Classical Physics}
\author{Jacob A. Barandes\thanks{Departments of Philosophy and Physics, Harvard University, Cambridge, MA 02138; jacob\_barandes@harvard.edu; ORCID: 0000-0002-3740-4418}
\thanks{This manuscript is the accepted version, as published in \emph{The European Physical Journal H} (DOI:10.1140/epjh/s13129-025-00113-x).}}
\date{}

\maketitle

\begin{abstract}
Hilbert-space techniques are widely used not only for quantum theory,
but also for classical physics. Two important examples are the Koopman\textendash von
Neumann (KvN) formulation and the method of ``classical'' wave functions.
As this paper explains, these two approaches are conceptually distinct.
In particular, the method of classical wave functions was not due
to Bernard Koopman and John von Neumann, but was developed independently
by a number of later researchers, perhaps first by Mario Schönberg,
with key contributions from Angelo Loinger, Giacomo Della Riccia,
Norbert Wiener, and E. C. George Sudarshan. The primary goals of this
paper are to explain these two approaches, describe the relevant history
in detail, and give credit where credit is due.
\end{abstract}

\begin{center}
\global\long\def\quote#1{``#1"}%
\global\long\def\apostrophe{\textrm{'}}%
\global\long\def\slot{\phantom{x}}%
\global\long\def\eval#1{\left.#1\right\vert }%
\global\long\def\keyeq#1{\boxed{#1}}%
\global\long\def\importanteq#1{\boxed{\boxed{#1}}}%
\global\long\def\given{\vert}%
\global\long\def\mapping#1#2#3{#1:#2\to#3}%
\global\long\def\composition{\circ}%
\global\long\def\set#1{\left\{  #1\right\}  }%
\global\long\def\setindexed#1#2{\left\{  #1\right\}  _{#2}}%

\global\long\def\setbuild#1#2{\left\{  \left.\!#1\,\right|\,#2\right\}  }%
\global\long\def\complem{\mathrm{c}}%

\global\long\def\union{\cup}%
\global\long\def\intersection{\cap}%
\global\long\def\cartesianprod{\times}%
\global\long\def\disjointunion{\sqcup}%

\global\long\def\isomorphic{\cong}%

\global\long\def\setsize#1{\left|#1\right|}%
\global\long\def\defeq{\equiv}%
\global\long\def\conj{\ast}%
\global\long\def\overconj#1{\overline{#1}}%
\global\long\def\re{\mathrm{Re\,}}%
\global\long\def\im{\mathrm{Im\,}}%

\global\long\def\transp{\mathrm{T}}%
\global\long\def\tr{\mathrm{tr}}%
\global\long\def\adj{\dagger}%
\global\long\def\diag#1{\mathrm{diag}\left(#1\right)}%
\global\long\def\dotprod{\cdot}%
\global\long\def\crossprod{\times}%
\global\long\def\Probability#1{\mathrm{Prob}\left(#1\right)}%
\global\long\def\Amplitude#1{\mathrm{Amp}\left(#1\right)}%
\global\long\def\cov{\mathrm{cov}}%
\global\long\def\corr{\mathrm{corr}}%

\global\long\def\absval#1{\left\vert #1\right\vert }%
\global\long\def\expectval#1{\left\langle #1\right\rangle }%
\global\long\def\op#1{\hat{#1}}%

\global\long\def\bra#1{\left\langle #1\right|}%
\global\long\def\ket#1{\left|#1\right\rangle }%
\global\long\def\braket#1#2{\left\langle \left.\!#1\right|#2\right\rangle }%

\global\long\def\parens#1{(#1)}%
\global\long\def\bigparens#1{\big(#1\big)}%
\global\long\def\Bigparens#1{\Big(#1\Big)}%
\global\long\def\biggparens#1{\bigg(#1\bigg)}%
\global\long\def\Biggparens#1{\Bigg(#1\Bigg)}%
\global\long\def\bracks#1{[#1]}%
\global\long\def\bigbracks#1{\big[#1\big]}%
\global\long\def\Bigbracks#1{\Big[#1\Big]}%
\global\long\def\biggbracks#1{\bigg[#1\bigg]}%
\global\long\def\Biggbracks#1{\Bigg[#1\Bigg]}%
\global\long\def\curlies#1{\{#1\}}%
\global\long\def\bigcurlies#1{\big\{#1\big\}}%
\global\long\def\Bigcurlies#1{\Big\{#1\Big\}}%
\global\long\def\biggcurlies#1{\bigg\{#1\bigg\}}%
\global\long\def\Biggcurlies#1{\Bigg\{#1\Bigg\}}%
\global\long\def\verts#1{\vert#1\vert}%
\global\long\def\bigverts#1{\big\vert#1\big\vert}%
\global\long\def\Bigverts#1{\Big\vert#1\Big\vert}%
\global\long\def\biggverts#1{\bigg\vert#1\bigg\vert}%
\global\long\def\Biggverts#1{\Bigg\vert#1\Bigg\vert}%
\global\long\def\Verts#1{\Vert#1\Vert}%
\global\long\def\bigVerts#1{\big\Vert#1\big\Vert}%
\global\long\def\BigVerts#1{\Big\Vert#1\Big\Vert}%
\global\long\def\biggVerts#1{\bigg\Vert#1\bigg\Vert}%
\global\long\def\BiggVerts#1{\Bigg\Vert#1\Bigg\Vert}%
\global\long\def\ket#1{\vert#1\rangle}%
\global\long\def\bigket#1{\big\vert#1\big\rangle}%
\global\long\def\Bigket#1{\Big\vert#1\Big\rangle}%
\global\long\def\biggket#1{\bigg\vert#1\bigg\rangle}%
\global\long\def\Biggket#1{\Bigg\vert#1\Bigg\rangle}%
\global\long\def\bra#1{\langle#1\vert}%
\global\long\def\bigbra#1{\big\langle#1\big\vert}%
\global\long\def\Bigbra#1{\Big\langle#1\Big\vert}%
\global\long\def\biggbra#1{\bigg\langle#1\bigg\vert}%
\global\long\def\Biggbra#1{\Bigg\langle#1\Bigg\vert}%
\global\long\def\braket#1#2{\langle#1\vert#2\rangle}%
\global\long\def\bigbraket#1#2{\big\langle#1\big\vert#2\big\rangle}%
\global\long\def\Bigbraket#1#2{\Big\langle#1\Big\vert#2\Big\rangle}%
\global\long\def\biggbraket#1#2{\bigg\langle#1\bigg\vert#2\bigg\rangle}%
\global\long\def\Biggbraket#1#2{\Bigg\langle#1\Bigg\vert#2\Bigg\rangle}%
\global\long\def\angs#1{\langle#1\rangle}%
\global\long\def\bigangs#1{\big\langle#1\big\rangle}%
\global\long\def\Bigangs#1{\Big\langle#1\Big\rangle}%
\global\long\def\biggangs#1{\bigg\langle#1\bigg\rangle}%
\global\long\def\Biggangs#1{\Bigg\langle#1\Bigg\rangle}%

\global\long\def\vec#1{\mathbf{#1}}%
\global\long\def\vecgreek#1{\boldsymbol{#1}}%
\global\long\def\idmatrix{\mymathbb{1}}%
\global\long\def\projector{P}%
\global\long\def\permutationmatrix{\Sigma}%
\global\long\def\densitymatrix{\rho}%
\global\long\def\krausmatrix{K}%
\global\long\def\stochasticmatrix{\Gamma}%
\global\long\def\lindbladmatrix{L}%
\global\long\def\dynop{\Theta}%
\global\long\def\timeevop{U}%
\global\long\def\hadamardprod{\odot}%
\global\long\def\tensorprod{\otimes}%

\global\long\def\inprod#1#2{\left\langle #1,#2\right\rangle }%
\global\long\def\normket#1{\left\Vert #1\right\Vert }%
\global\long\def\hilbspace{\mathcal{H}}%
\global\long\def\samplespace{\Omega}%
\global\long\def\configspace{\mathcal{C}}%
\global\long\def\phasespace{\mathcal{P}}%
\global\long\def\spectrum{\sigma}%
\global\long\def\restrict#1#2{\left.#1\right\vert _{#2}}%
\global\long\def\from{\leftarrow}%
\global\long\def\statemap{\omega}%
\global\long\def\degangle#1{#1^{\circ}}%
\global\long\def\trivialvector{\tilde{v}}%
\global\long\def\eqsbrace#1{\left.#1\qquad\right\}  }%
\par\end{center}

\section{Introduction\label{sec:Introduction}}

In 1931, Bernard Koopman published a paper titled ``Hamiltonian Systems
and Transformations in Hilbert Space'' in \emph{Proceedings of the
National Academy of Sciences} (Koopman 1931)\nocite{Koopman:1931hsatihs}.
Koopman's paper laid out a novel method for identifying functions
representing observables on a classical system's phase space as vectors
in a new kind of Hilbert space. In 1932, John von Neumann published
a pair of follow-up papers in German in \emph{Annals of Mathematics}
(von Neumann 1932a, 1932b)\nocite{vonNeumann:1932zoidkm,vonNeumann:1932zzazo}
further developing Koopman's method.

Koopman's use of Greek letters and complex values for his phase-space
functions may have made it easy to confuse them with classical versions
of the state vectors or wave functions of quantum mechanics, and von
Neumann's two papers were never translated into English. Koopman and
von Neumann's papers, however, did not refer to classical state vectors
or wave functions, and their Hilbert spaces consisted of functions
representing classical observables, which naturally corresponded to
quantum-mechanical operators evolving in time in the Heisenberg picture.
Indeed, as Jordan and Sudarshan noted in a 1961 paper published in
\emph{Journal of Mathematical Physics}:
\begin{quotation}
It was shown by Koopman how the dynamical transformations of classical
mechanics, considered as measure preserving transformations of the
phase space, induce unitary transformations on the Hilbert space of
functions which are square integrable with respect to a density function
over the phase space. This Hilbert space formulation of classical
mechanics was further developed by von Neumann. It is to be noted
that this Hilbert space corresponds not to the space of state vectors
in quantum mechanics but to the Hilbert space of operators on the
state vectors (with the trace of the product of two operators being
chosen as the scalar product). {[}Jordan, Sudarshan 1961, pp. 515\textendash 516{]}\nocite{JordanSudarshan:1961lgdfatrbqmacm}
\end{quotation}

Danilo Mauro wrote an innovative and influential 2002 paper titled
``On Koopman\textendash von Neumann Waves'' (Mauro 2002)\nocite{Mauro:2002oknw},
later expanding on his work in his 2003 PhD thesis, titled ``Topics
in Koopman\textendash von Neumann Theory'' (Mauro 2003)\nocite{Mauro:2003tikvnt}.
The paper and thesis replaced the observables-as-vectors method described
above with an important but different technique. This different technique
was to use complex-valued ``classical'' wave functions $\psi$ in
place of classical probability distributions $\rho$, where these
classical wave functions and classical probability distributions were
explicitly related by the modulus-square operation, $\rho=\verts{\psi}^{2}$,
in analogy with the Born rule. However, this method of classical wave
functions was due to other researchers who came decades after Koopman
and von Neumann's papers from the 1930s\textemdash perhaps the first
being Mario Schönberg, working in the 1950s.

Ever since this accidental misattribution, the ``Koopman\textendash von
Neumann (KvN) formulation'' has been widely but incorrectly employed
to refer to the method of classical wave functions. For example, an
international conference in 2021 and an accompanying special issue
of \emph{Journal of Physics A} in 2022, both titled ``Koopman Methods
in Classical and Classical-Quantum Mechanics,'' referred in their
abstracts to ``Koopman\textendash von Neumann wave functions'' (Bondar
et al., 2021, 2022)\nocite{BondarBurghardtGay-BalmazMezicTronci:2021kmicacqm,BondarBurghardtGay-BalmazMezicTronci:2022kmicaqcm}.
As of the writing of this paper, the Wikipedia entry ``Koopman\textendash von
Neumann Classical Mechanics'' (Wikipedia 2025)\nocite{Wikipediacontributors:2025kvncm}
opens up its derivation of the framework with the following statement: 
\begin{quotation}
In the approach of Koopman and von Neumann (KvN), dynamics in phase
space is described by a (classical) probability density, recovered
from an underlying wavefunction\textemdash the Koopman\textendash von
Neumann wavefunction\textemdash as the square of its absolute value
(more precisely, as the amplitude multiplied with its own complex
conjugate).
\end{quotation}
The main purpose of the present paper is to lay out the detailed history
of both the Koopman\textendash von Neumann formulation and the method
of classical wave functions, and assign credit appropriately.

Ultimately, it turns out that the Hilbert spaces that arise from treating
observables as vectors and the Hilbert spaces that arise from classical
wave functions are mathematically equivalent. This equivalence between
the two kinds of Hilbert spaces is an elementary result of the GNS
construction (Gelfand, Naimark 1943; Segal 1947)\nocite{GelfandNaimark:1943otionritroooahs,Segal:1947otionritroooahs},
which takes the elements $f,g,\dots$ of a C{*}-algebra (representing
observables) together with a positive, normalized linear functional
$\omega$ in the dual space of the C{*}-algebra (representing a quantum
state), and combines them to form a rudimentary inner product $\left(f,g\right)\defeq\omega\left(f^{\star}g\right)$
that eventually underwrites the definition of a Hilbert space representing
the original C{*}-algebra. However, despite this underlying connection
between, on the one hand, the original observables-as-vectors method
of Koopman and von Neumann, and, on the other hand, the classical-wave-function
method that came later, these are conceptually different methods.
Simply put, Koopman and von Neumann did not come up with the idea
of using classical wave functions to capture classical probability
distributions.

By analogy, Heisenberg's matrix mechanics (Heisenberg 1925)\nocite{Heisenberg:1925uqukumb}
and Schrödinger's wave mechanics (Schrödinger 1926)\nocite{Schrodinger:1926autotmoaam}
were conceptually different frameworks. It was Schrödinger who came
up with the idea of quantum-mechanical wave functions, and it would
not be correct to give credit to Heisenberg for that idea, even though
matrix mechanics and wave mechanics were eventually connected to each
other by modern Hilbert-space formulations of quantum theory.

\section{Hilbert-Space Formulations of Classical Physics\label{sec:Hilbert-Space-Formulations-of-Classical-Physics}}

\subsection{Bernard Koopman\label{subsec:Bernard-Koopman}}

Koopman began his 1931 paper with the following motivation: 
\begin{quotation}
In recent years the theory of Hilbert space and its linear transformations
has come into prominence. {[}...{]} It is the object of this note
to outline certain investigations of our own in which the domain of
this theory has been extended in such a way as to include classical
Hamiltonian mechanics, or, more generally, systems defining a steady
$n$-dimensional flow of a fluid of positive density. {[}Koopman 1931,
p. 315{]}\nocite{Koopman:1931hsatihs}
\end{quotation}
Shortly thereafter, Koopman described his basic approach: 
\begin{quotation}
The starting point of our investigation is the $N$-dimensional variety
$\Omega$ and the group of automorphisms $S_{t}$ having the positive
integral invariant $\int\rho\,d\omega$, and these are considered
without reference to the problem which gave them origin. Let $\varphi=\varphi\left(A\right)$
be a complex-valued function of the point $A$ of $\Omega$, restricted
only as follows: (i) $\varphi$ is single-valued; (ii) $\varphi$
is measurable; (iii) the Lebesgue integrals $\int_{\Omega}\rho\verts{\varphi}d\omega$
and $\int_{\Omega}\rho\verts{\varphi}^{2}d\omega$ are finite. The
totality of such functions $\varphi$ constitutes the aggregate of
points of a Hilbert space $\mathfrak{H}$: the metric of which is
determined by the ``inner product'' 
\begin{equation}
\left(\varphi,\psi\right)=\int_{\Omega}\rho\varphi\bar{\psi}d\omega.\label{eq:DefKoopmanInnerProduct}
\end{equation}
 {[}Ibid., p. 316{]}
\end{quotation}
Notice the function $\rho$ appearing in Koopman's integral measures,
separate from the functions $\varphi$ and $\psi$. As Koopman wrote,
``here, $\rho$ is a positive, single-valued, analytic function on
$\Omega$. This is a consequence of the fact that $\int dq_{1}\dots dq_{n}dp_{1}\dots dp_{n}$
is an integral invariant of the system'' (Ibid., p. 315). Later,
Koopman added: ``If $t$ represents the time, $S_{t}$ specifies
the steady flow of a fluid of density $\rho$ occupying the space
$\Omega$'' (Ibid., p. 316). After introducing Gaussian coordinates
$\xi_{1},\dots,\xi_{N}$, with corresponding velocities $\Xi_{k}=d\xi_{k}/dt$,
Koopman wrote: 
\begin{quotation}
The property of $\rho=\rho\left(\xi_{1},\dots,\xi_{N}\right)$ is
expressed by the ``equation of continuity'' 
\begin{equation}
\sum^{N}_{k=1}\frac{\partial\left(\rho\Xi_{k}\right)}{\partial\xi_{k}}=0.\label{eq:KoopmanContinuityEq}
\end{equation}
\end{quotation}
These statements make clear that Koopman's functions $\varphi$ and
$\psi$ were not related to probability densities, and that his density
function $\rho$ was a separate mathematical object that partly defined
Koopman's Hilbert space. Koopman did not suggest that his functions
$\varphi$ and $\psi$ should include classical wave functions, or
be related to a probability density by the modulus-square operation.

In Koopman's paper, he defined the time evolution of his functions
$\varphi,\psi,\dots$ using a transformation $U_{t}$ defined by 
\begin{equation}
U_{t}\varphi\left(A\right)\defeq\varphi\left(S_{t}A\right),\label{eq:DefKoopmanTimeEv}
\end{equation}
 where $A$ is a given phase-space point and where $S_{t}A$ is the
new phase-space point after a duration $t$ of classical Hamiltonian
time evolution. Working in coordinates for the given $2n$-dimensional
phase space, so that one can denote a phase-space point as $A=\left(q,p\right)\defeq\left(q_{1},\dots,q_{n};p_{1},\dots,p_{n}\right)$,
one has a coordinate representation of Koopman's notion of time evolution,
\begin{equation}
S_{t}\left(q,p\right)\defeq\left(q\left(t\right),p\left(t\right)\right),\label{eq:DefKoopmanTimeEvInCoordinates}
\end{equation}
 and so $U_{t}\varphi\left(A\right)$ has time derivative given by
\begin{align*}
\frac{d}{dt}U_{t}\varphi\left(A\right) & =\frac{d}{dt}\varphi\left(S_{t}A\right)\\
 & =\frac{d}{dt}\varphi\left(q\left(t\right),p\left(t\right)\right)\\
 & =\sum^{n}_{k=1}\frac{\partial\varphi}{\partial q_{k}}\frac{dq_{k}\left(t\right)}{dt}+\sum^{n}_{k=1}\frac{\partial\varphi}{\partial p_{k}}\frac{dp_{k}\left(t\right)}{dt}\\
 & =\sum^{n}_{k=1}\left(\frac{\partial\varphi}{\partial q_{k}}\frac{\partial H}{\partial p_{k}}-\frac{\partial\varphi}{\partial p_{k}}\frac{\partial H}{\partial q_{k}}\right)\\
 & =\left\{ \varphi,H\right\} ,
\end{align*}
 which is the appropriate time-evolution equation for a function representing
a classical observable, with $\left\{ \varphi,H\right\} $ the usual
Poisson bracket of $\varphi$ and $H$. Introducing the standard formula
for the Liouvillian operator $L$, 
\begin{equation}
L\defeq i\left\{ H,\slot\right\} \defeq i\sum^{n}_{k=1}\left(\frac{\partial H}{\partial q_{k}}\frac{\partial}{\partial p_{k}}-\frac{\partial H}{\partial p_{k}}\frac{\partial}{\partial q_{k}}\right),\label{eq:DefLiouvillianOp}
\end{equation}
 one can recast the time-evolution equation for $\varphi$ as\footnote{Koopman wrote in his paper that ``if the values of $\varphi\left(A\right)$
be regarded as being attached to the respective points $A$ of the
fluid when $t=0$, in the course of the flow these values will be
carried into those of the function $U_{-t}\varphi\left(A\right)$''
(Koopman 1931, p. 316). This picture led Koopman to calculate instead
$\left[\left(\partial/\partial t\right)U_{t}\varphi\left(A\right)\right]_{t=0}=iP\varphi\left(A\right)$,
where $P$ differed by an overall sign from the usual definition of
the Liouvillian operator $L$ in \eqref{eq:DefLiouvillianOp}. Koopman's
equation, however, was not a time-evolution equation in the usual
sense of describing the behavior of a given function at arbitrary
times $t$. Instead, Koopman treated his equation merely as a mathematical
step toward writing down a self-adjoint generator $P$ for his transformation
$U_{t}$.} 
\begin{equation}
\frac{d}{dt}U_{t}\varphi\left(A\right)=\left\{ \varphi,H\right\} =iL\varphi.\label{eq:KoopmanFunctionTimeEvEqFromLiouvillianOp}
\end{equation}
 By contrast, under the time evolution $S_{t}\left(q,p\right)\defeq\left(q\left(t\right),p\left(t\right)\right)$
expressed in \eqref{eq:DefKoopmanTimeEvInCoordinates}, a time-dependent
probability density $\rho\left(q,p,t\right)$ on the classical phase
space should evolve instead according to the classical Liouville equation:
\begin{equation}
\frac{\partial\rho}{\partial t}=\left\{ H,\rho\right\} =-iL\rho.\label{eq:LiouvilleEq}
\end{equation}
 Because the Liouvillian operator \eqref{eq:DefLiouvillianOp} involves
only first-order derivatives, the same should be true of any classical
wave function $\psi$ whose modulus-square is $\rho$.

Finally, for the case of a probability density $\rho$ without any
explicit time-dependence, the classical Liouville equation reduces
to 
\[
\left\{ H,\rho\right\} =0,
\]
 which implies that 
\[
\sum^{n}_{k=1}\frac{\partial\rho}{\partial q_{k}}\frac{dq_{k}\left(t\right)}{dt}+\sum^{n}_{k=1}\frac{\partial\rho}{\partial p_{k}}\frac{dp_{k}\left(t\right)}{dt}=0,
\]
 exactly in keeping with Koopman's continuity equation \eqref{eq:KoopmanContinuityEq}.
These results confirm that Koopman's function $\rho$ really does
correspond most closely with a classical system's probability density,
and is conceptually distinct from Koopman's phase-space functions
$\varphi,\psi,\dots$.

From Koopman's notion of time evolution, \eqref{eq:DefKoopmanTimeEv},
one can see at an even deeper level why his phase-space functions
were not akin to wave functions. For any classical phase-space point
$A$, let $\omega_{A}$ be a positive linear functional acting on
Koopman's functions $\varphi,\psi,\dots$ according to 
\begin{equation}
\omega_{A}\left(\varphi\right)\defeq\varphi\left(A\right),\label{eq:DefGNSStateOnKoopmanFunction}
\end{equation}
 and satisfying the usual desiderata of a state map in the C{*}-algebraic
sense. For any time $t$, let $g_{t}$ be the map acting on $\omega_{A}$
by replacing $A$ with its time-evolved counterpart $S_{t}A$, in
accordance with Koopman's notion of time evolution: 
\begin{equation}
g_{t}\omega_{A}\defeq\omega_{S_{t}A}.\label{eq:DefGNSStateTimeEvForKoopman}
\end{equation}
 It follows immediately that 
\begin{equation}
\left(g_{t}\omega_{A}\right)\left(\varphi\right)=\omega_{A}\left(U_{t}\varphi\right).\label{eq:GNSTimeEvolutionOnEvaluatedStateMap}
\end{equation}
 Evidently, the time evolution of state maps $\omega_{A}$ is opposite
to the time evolution of functions $\varphi$, in an abstraction of
the usual distinction between Schrödinger-picture time evolution and
Heisenberg-picture time evolution. According to Koopman's transformation
$U_{t}$, his phase-space functions time-evolve as Heisenberg-picture
observables, not as Schrödinger-picture wave functions.

\subsection{John von Neumann\label{subsec:John-von-Neumann}}

Koopman did not mention wave functions at all in his paper. By contrast,
von Neumann did mention wave functions, but only in the first of his
two 1932 papers (von Neumann 1932a)\nocite{vonNeumann:1932zoidkm},
and only to suggest an analogy between Koopman's time-evolution operator
$U_{t}$ and the time evolution appearing in quantum mechanics:
\begin{quotation}
Finally, we would like to point out the interesting analogy between
Koopman\textquoteright s operators $U_{t}=e^{itA}$ and the operators
of quantum mechanics. The Schrödinger wave function $\varphi$ (defined
in the state space of the mechanical system and not like our $f$
in phase space!) obeys, as is well known, in its dependence on the
time parameter $t$ the differential equation $\frac{h}{2\pi i}\frac{\partial\varphi}{\partial t}=H\varphi$.
Here $h$ is Planck\textquoteright s quantum of action and $H$ the
energy operator. From this it follows at once $\varphi=e^{it\cdot\frac{2\pi}{h}H}\varphi_{\left(t=0\right)}$
{[}footnote in the original: This relationship is usually written
with $-H$ instead of $H$.{]}, so that here the unitary operators
$\hat{U}_{t}=e^{it\cdot\frac{2\pi}{h}H}$ play a fundamental role.
The analogy, which arises from the juxtaposition of $A$ and ${\displaystyle \frac{2\pi}{h}H}$,
is striking {[}footnote in the original: A closer look shows that
it becomes more perfect if one replaces the differential equation
of the wave function with that of the so-called statistical operator
(cf., for example, J. v. Neumann, \emph{Mathematische Grundlagen der
Quantenmechanik}, Berlin 1932, p. 186). However, it is constructed
in the same way, and what will be said below also applies to it.{]},
and it is even possible to exhibit the continuous passage of quantum
mechanics into the classical (as $h\to0$). Nevertheless, there seem
to be essential mathematical differences between these two families
of operators. For a mechanical system confined to a finite volume,
quantum mechanics always appears to exhibit a pure point spectrum,
whereas in the classical-mechanical problem a pure continuous spectrum
seems to be the generic case (cf. § VI). {[}Ibid., pp. 594\textendash 595{]}\nocite{vonNeumann:1932zoidkm}\footnote{In the original German:
\begin{quotation}
Zum Schluß sei noch auf die interessante Analogie zwischen Koopmans
Operatoren $U_{t}=e^{itA}$ und den Operatoren der Quantenmechanik
hingewiesen. Die Schrödingersche Wellenfunktion $\varphi$ (definiert
im Zustandsraume des mechanischen Systems und nicht wie unsere $f$
im Phasenraume!) gehorchtbekanntlich in ihrer Abhängigkeit vom Zeitparameter
t der Differentialgleichung $\frac{h}{2\pi i}\frac{\partial}{\partial t}\varphi=H\varphi$.
Hier ist $h$ das Plancksche Wirkungsquantum under $H$ der Energieoperator.
Hieraus folgt sofort $\varphi=e^{it\cdot\frac{2\pi}{h}H}\varphi_{\left(t=0\right)}$
{[}Meistens wird diese Beziehung mit -H statt H geschrieben{]}, so
daß hier die unitären Operatoren $\hat{U}_{t}=e^{it\cdot\frac{2\pi}{h}H}$
eine fundamentale Rolle spielen. Die Analogie, die durch das Nebeneinanderstellen
von $A$ und $\frac{2\pi}{h}H$ entsteht, ist auffallend {[}Eine genauere
Überlegung zeigt, daß sie vollkommener wird, wenn man die Differentialgleichung
der Wellenfunktion durch diejenige des sog. statistischen Operators
ersetzt (vgl. z. B. J. v. Neumann, Mathematische Grundlagen der Quantenmechanik,
Berlin 1932, S. 186). Dieselbe ist aber ebenso gebaut und das weiter
unten zu Sagende gilt auch für sie.{]}, und es ist möglich, sie zum
Nachweis des stetigen Übergehens der Quantenmechanik in die klassische
(für $h\to0$) auszubauen. Trotzdem scheinen wesentliche mathematische
Unterschiede zwischen diesen Operatorenscharen zu bestehen. Denn für
ein mechanisches System, das in ein endliches Volumen eingesperrt
ist, scheint in der Quantenmechanik stets ein reines Punktspektrum
vorzuliegen, während im klassisch-mechanischen Problem das reine Streckenspektrum
der allgemeine Fall zu sein scheint (vgl. $\S$ VI).
\end{quotation}
}
\end{quotation}
Notice the footnote in which von Neumann pointed out that his phase-space
function $f$ should evolve under $-H$ rather than $H$. Once again,
this reversed time evolution is due to $f$ being akin to an observable,
and not a wave function.

\subsection{Mario Schönberg\label{subsec:Mario-Sch=0000F6nberg}}

In the early 1950s, Mario Schönberg published a pair of papers in
\emph{Il Nuovo Cimento} (Schönberg 1952, 1953)\nocite{Schonberg:1952aosqmttcsm,Schonberg:1953aosqmttcsmi}
in which he introduced the method of classical wave functions, an
idea that did not appear in the 1930s papers by Koopman or von Neumann.
Again, a classical wave function is a complex-valued function whose
modulus-square gives the probability density for a classical system. 

Following Schönberg and his notation, he considered a system of $n$
particles with positions $\vec x_{1},\dots,\vec x_{n}$ and momenta
$\vec p_{1},\dots,\vec p_{n}$, as well as a probability density $f_{n}$.
Schönberg called his new complex-valued function $\Theta_{n}$, and
then his equation (16) from his 1952 paper took the form: 
\begin{equation}
f_{n}\left(\vec x_{1},\dots,\vec x_{n};\vec p_{1},\dots,\vec p_{n}\right)=\verts{\Theta_{n}\left(\vec x_{1},\dots,\vec x_{n};\vec p_{1},\dots,\vec p_{n}\right)}^{2}\qquad\left[\textrm{Schönberg's eq. }\left(16\right)\right].\label{eq:SchonbergProbatilityDensityFromModSquare}
\end{equation}
 Earlier in his paper, in his equation (8), Schönberg had noted that
the probability density $f_{n}$ obeyed the classical Liouville equation,
in accord with \eqref{eq:LiouvilleEq}, 
\begin{equation}
\frac{\partial f_{n}}{\partial t}=\left(H_{n},f_{n}\right)_{n}=-iL_{n}f_{n}\qquad\left[\textrm{Schönberg's eq. }\left(8\right)\right],\label{eq:SchonbergLiouvilleEq}
\end{equation}
 where $L_{n}$ is the $n$-particle Liouville operator, and where
Schönberg used the following notation for Poisson brackets: 
\begin{equation}
\left(F,G\right)_{n}=\sum^{n}_{l=1}\left\{ \frac{\partial F}{\partial\vec x_{l}}\frac{\partial G}{\partial\vec p_{l}}-\frac{\partial F}{\partial\vec p_{l}}\frac{\partial G}{\partial\vec x_{l}}\right\} \qquad\left[\textrm{Schönberg's eq. }\left(9\right)\right].\label{eq:SchonbergPoissonBrackets}
\end{equation}
 As Schönberg then explained, ``since the square of the absolute
value of a solution of the Liouville equation is also a solution of
the same equation,'' it followed that his new complex-valued function
$\Theta_{n}$ satisfied the same equation: 
\begin{equation}
\frac{\partial\Theta_{n}}{\partial t}=\left(H_{n},\Theta_{n}\right)_{n}=-iL_{n}\Theta_{n}\qquad\left[\textrm{Schönberg's eq. }\left(17\right)\right].\label{eq:SchonbergWaveFunctionLiouvilleEq}
\end{equation}
 In the paragraph that followed this equation, Schönberg wrote: 
\begin{quotation}
Thus we are led to a kind of wave function in classical statistical
mechanics. Equation (17) may be considered as the classical wave equation,
the {[}H{]}ermitian operator $L_{n}$ playing the part of {[}a{]}
classical {[}H{]}amiltonian operator. {[}Schönberg 1952, p. 1142{]}\nocite{Schonberg:1952aosqmttcsm}
\end{quotation}

In the opening of his 1953 paper, Schönberg wrote:
\begin{quotation}
In the preceding part of {[}Schönberg 1952{]} we have shown that it
is possible to develop in the classical mechanics a wave formalism
in phase space which presents many of the features of the quantum
wave mechanics. {[}...{]} {[}W{]}e may introduce a wave function $\Theta\left(q,p\right)$,
in general complex, such that the probability density be the square
of its absolute value: 
\begin{equation}
f\left(q,p\right)=\verts{\Theta\left(q,p\right)}^{2}\qquad\left[\textrm{Schönberg's eq. }\left(3\right)\right].\label{eq:SchoenbergSecondPaperProbabilityDensityFromModSquare}
\end{equation}
 The consideration of the wave function gives some essential new possibilities,
because it is not restricted to have only real and positive values,
as the probability density. {[}Schönberg 1953, p. 419{]}\nocite{Schonberg:1953aosqmttcsmi}
\end{quotation}

In that 1953 paper, Schönberg made only a single reference to Koopman's
1931 paper, pointing to the underlying mathematical equivalence between
Schönberg's Hilbert spaces and Koopman's Hilbert spaces that was described
in Section~\ref{sec:Introduction} of the present paper: 
\begin{quotation}
The introduction of the classical functions clarifies the meaning
of the unitary transformations in Hilbert space associated with the
motion of a classical system, which were introduced by Koopman. The
Koopman Hilbert-space can be taken as that of the classical wave functions.
{[}Schönberg 1953, p. 425{]}
\end{quotation}
It is unclear whether Schönberg meant something more by these remarks,
such as asserting not just a correspondence between underlying Hilbert
spaces, but a closer relationship between his classical wave functions
and Koopman's phase-space functions. The answer may be lost to history.

\subsection{Angelo Loinger\label{subsec:Angelo-Loinger}}

In 1962, Angelo Loinger published a paper titled ``Galilei Group
and Liouville Equation'' in \emph{Annals of Physics} (Loinger 1962)\nocite{Loinger:1962ggale}.
The paper explicitly distinguished between the original Koopman\textendash von
Neumann construction and Schönberg's work. In his introduction, Loinger
wrote: 
\begin{quotation}
In Section I the Hilbert space formulation of classical mechanics,
according to the viewpoints developed respectively by Koopman and
by Schönberg, is briefly reviewed in a convenient form and the intimate
relationship existing between the two conceptions is pointed out.
{[}Ibid., p. 132{]}\nocite{Loinger:1962ggale}
\end{quotation}

Loinger then went on to review the original Koopman\textendash von
Neumann approach in Section I.A. of the paper (``Preliminaries and
Koopman's Viewpoint''). In that section, Loinger slightly altered
the definition of the inner product \eqref{eq:DefKoopmanInnerProduct},
writing: 
\begin{quotation}
We then consider the Hilbert space $\mathfrak{K}$ of the complex
square integrable functions $f\left(\omega\right)$ of the points
of $\Omega_{2n}$ {[}the classical system's $2n$-dimensional phase
space{]}, in which an inner product is defined as follows: 
\begin{equation}
\left(f,g\right)\overset{\textrm{Def.}}{=}\int_{\Omega}f^{\ast}\left(\omega\right)g\left(\omega\right)\,d\omega\qquad\left[\textrm{Loinger's eq. }\left(9\right)\right].\label{eq:DefLoingerInnerProduct}
\end{equation}
 {[}Ibid., p. 134{]}
\end{quotation}
Note the absence of the phase-space probability density $\rho$ in
this definition, as compared with Koopman's original definition \eqref{eq:DefKoopmanInnerProduct}.
Loinger then wrote: 
\begin{quotation}
We put, with a bra and ket notation: 
\begin{equation}
\left(f,g\right)\defeq\left(f\vert g\right)\qquad\left[\textrm{Loinger's eq. }\left(11\right)\right]\label{eq:DefLoingerDiracBraKet}
\end{equation}
 and consider the extension $\mathfrak{K}_{D}$ of $\mathfrak{K}$
in Dirac's sense. {[}Ibid., p. 134{]}
\end{quotation}

Next, Loinger turned to a review of Schönberg's framework in Section
II.A (``Schönberg's Viewpoint''), writing:
\begin{quotation}
The formulation of the classical mechanics given in Section I, A is
formally analogous to a formulation of quantum mechanics, which was
first studied by von Neumann. In this von Neumann quantal scheme the
operators of the conventional formulation are considered as vectors
of a new Hilbert space. One may ask whether even in the classical
case it is possible to introduce a Hilbert space, say $\mathfrak{H}_{D}$,
analogous to the state-vector space of quantum mechanics. It can easily
be seen that such a space exists and stays in a very simple relationship
with $\mathfrak{K}_{D}$. {[}Ibid., pp. 134\textendash 135{]}
\end{quotation}
Loinger added parenthetically that he would show later that the two
Hilbert spaces were mathematically equivalent, writing: 
\begin{quotation}
(Actually, as will be apparent presently, $\mathfrak{H}_{D}$ and
$\mathfrak{K}_{D}$ are the same space.) {[}Ibid., p. 135{]}
\end{quotation}
However, throughout the paper, Loinger emphasized that despite the
isomorphism between $\mathfrak{H}_{D}$ and $\mathfrak{K}_{D}$, they
had conceptually different meanings. For example, he included the
following additional parenthetical note:
\begin{quotation}
(We distinguish here the vectors of $\mathfrak{K}_{D}$ from those
of $\mathfrak{H}_{D}$, reserving for the first ones the bra and ket
notation with round parentheses.) {[}Ibid., p. 135{]}
\end{quotation}

\subsection{Giacomo Della Riccia and Norbert Wiener\label{subsec:Giacomo-Della-Riccia-and-Norbert-Wiener}}

In 1966, \emph{Journal of Mathematical Physics} published a paper
by Giacomo Della Riccia and (posthumously)\footnote{The paper included the following note, at the end of its introductory
section: ``Because of the sad demise of Norbert Wiener in March 1964,
the treatment given in this paper is due to the first-named author.
For the same reason it seemed desirable that the results should be
presented, however incomplete they may be.''} Norbert Wiener (Della Riccia, Wiener 1966)\nocite{DellaRicciaWiener:1966wmicpsbmaqt},
titled ``Wave Mechanics in Classical Phase Space, Brownian Motion,
and Quantum Theory.'' In their abstract, the authors wrote:
\begin{quotation}
A wave dynamics of fields $\varphi\left(p,q;t\right)\in L_{2}\left(\Gamma\right)$
over the phase space $\Gamma\left(p,q\right)$ of a classical system
$\mathcal{S}$ is derived from the Liouville theorem. {[}...{]} From
this it follows that we can regard normalized fields $\varphi\left(p,q;t\right)$
as ``probability amplitudes'' leading to a probability density function
$\rho\left(p,q;t\right)=\varphi\varphi^{\ast}$ in the sense of Gibbs'
statistical mechanics. {[}Ibid., p. 1372{]}
\end{quotation}
In their introductory material, Della Riccia and Wiener wrote:
\begin{quotation}
In Gibbs statistical mechanics the basic quantity is a probability
density function $\rho\left(p,q;t\right)$ defined over the phase
space of the mechanical system $\mathcal{S}$ under observation. In
our work we introduce in phase space a new quantity $\varphi\left(p,q;t\right)$,
which by definition is a normalized square-integrable, real, or complex-valued
function. We call it a ``probability amplitude'' field. $\varphi\left(p,q;t\right)$
is required to satisfy the usual equation of continuity derived from
the Liouville theorem.

{[}...{]}

In the last part of our work we are concerned with the problem of
constructing probabilities out of ``probability amplitudes.'' We
use known results based on Wiener's mathematical theory of Brownian
motion to derive probabilities which agree with those obtained from
Born's statistical postulate. The desired result is that it is possible
to interpret the quantity $\rho\left(p,q;t\right)=\varphi\varphi^{\ast}$
as a probability density in the sense of Gibbs. {[}Ibid., pp. 1372\textendash 1373{]}
\end{quotation}
Della Riccia and Wiener made no mention of the 1930s papers by Koopman
or von Neumann, nor did they cite Schönberg or Loinger. It may be
that they did not see a connection between their classical probability
amplitudes and the Koopman\textendash von Neumann formulation, and
might not have been aware of Schönberg and Loinger's papers. Nevertheless,
Della Riccia and Wiener recapitulated Schönberg's derivation of the
time-evolution equation \eqref{eq:SchonbergWaveFunctionLiouvilleEq},
which they wrote as 
\begin{equation}
-i\left(\partial\varphi/\partial t\right)=\mathcal{L}\varphi,\qquad\varphi\in L_{2}\qquad\left[\textrm{Della Riccia and Wiener's eq. }\left(4\right)\right],\label{eq:DellaRicciaWienerWaveFunctionLiouvilleEq}
\end{equation}
 with their Liouville operator defined, as usual, according to $\mathcal{L}=i\left[H,\cdot\right]$,
where the right-hand side is Della Riccia and Wiener's notation for
the Poisson bracket with the Hamiltonian $H$.

\subsection{E. C. George Sudarshan\label{subsec:E.-C.-George-Sudarshan}}

In 1976, E. C. George Sudarshan laid out a very similar method in
a paper published in the journal \emph{Pramana}, titled ``Interaction
between Classical and Quantum Systems and the Measurement of Quantum
Observables'' (Sudarshan 1976)\nocite{Sudarshan:1976ibcaqsatmoqo}.
In the paper, Sudarshan described his motivation as trying to find
a better way to capture how classical systems and quantum systems
could interact with each other, especially during a measurement process.
In Sudarshan's own words:
\begin{quotation}
I introduce a direct method of dealing with the interaction of classical
and quantum systems. It is made possible by the discovery that a classical
system can be embedded in a quantum system with a continuum of superselection
sectors. {[}Ibid., p. 118{]}
\end{quotation}
In the second section of his paper, in defining a ``quantum system''
with this continuum of superselection sectors, whose eventual purpose
was to represent a classical system with canonical coordinates $\omega=\left(x,p\right)$,
Sudarshan wrote:
\begin{quotation}
State vectors for the quantum system are given, in the Schrödinger
representation, by their wave functions $\psi\left(\omega\right)$.
But because of the superselection principle, the relative phase of
the distinct ideal eigenstates of coordinate operators is unmeasurable
and, therefore, irrelevant. Hence, we are led to the equivalence 
\begin{equation}
\psi\left(\omega\right)\sim\psi\left(\omega\right)\exp\left\{ i\phi\left(\omega\right)\right\} \qquad\left[\textrm{Sudarshan's eq. }\left(2.8\right)\right].\label{eq:SudarshanWaveFunctionPhaseEquivalence}
\end{equation}
 Therefore, only the absolute value of $\psi\left(\omega\right)$
is relevant and may be taken as the positive square root of the phase
space density 
\begin{equation}
\psi\left(\omega\right)=\sqrt{\rho\left(\omega\right)}\qquad\left[\textrm{Sudarshan's eq. }\left(2.9\right)\right].\label{eq:SudarshanWaveFunctionAsSquareRootProbabilityDensity}
\end{equation}
 The ideal eigenstates of the coordinate operators is {[}sic{]} identified
with the classical state corresponding to a point in phase space.
{[}Ibid., p. 120{]}
\end{quotation}
Crucially, Sudarshan included the following bracketed note:
\begin{quotation}
{[}This construction of a quantum theory embedding the classical theory
is to be contrasted with the work of Coopman {[}sic{]} 1931; see also,
Jordan and Sudarshan 1961{]}. {[}Ibid., p. 120{]}
\end{quotation}
Sudarshan did not cite von Neumann, Schönberg, Loinger, Della Riccia,
or Wiener in his paper. In a later paper that further developed these
methods, published in \emph{Physical Review D} in 1978, Sudarshan
and his co-author, Tom Sherry, left out those names again, and did
not cite Koopman, either (Sherry, Sudarshan 1978)\nocite{SherrySudarshan:1978ibcaqsanatqmi}.
Sudarshan was clearly familiar with von Neumann's work, as the first
of von Neumann's 1932 papers was cited in Sudarshan's earlier 1961
paper with Jordan (Jordan, Sudarshan 1961)\nocite{JordanSudarshan:1961lgdfatrbqmacm},
as quoted in Section~\ref{sec:Introduction} of the present paper.
Perhaps the other papers were simply unknown to Sudarshan at the time. 

\subsection{Ennio Gozzi\label{subsec:Ennio-Gozzi}}

Danilo Mauro's PhD supervisor at the University of Trieste was Ennio
Gozzi. Gozzi had been working on formal connections between classical
mechanics and quantum theory as far back as 1988, when he published
a paper in \emph{Physical Letters B}, titled ``Hidden BRS Invariance
in Classical Mechanics,'' in which he derived a path-integral representation
of classical mechanics (Gozzi 1988)\nocite{Gozzi:1988hbiicm}. In
that paper, Gozzi also introduced a classical probability distribution
on phase space, writing, in a footnote, ``I owe this idea to a crucial
discussion with E. {[}Erhard{]} Seiler.'' As Gozzi wrote:
\begin{quotation}
This classical path-integral formalism has, as the quantum one, a
parallel operatorial version that is already well known. It is the
Liouville version of classical mechanics in which one introduces a
classical probability density $\rho\left(p,q\right)$ which evolves
in time as any other classical observables 
\begin{equation}
\frac{\partial\rho}{\partial t}=\left\{ \rho,H\right\} _{PB},\label{eq:GozziLiouvilleEq}
\end{equation}
 where $\left\{ P,B\right\} _{PB}$ are the usual Poisson brackets
and $H$ the {[}H{]}amiltonian. This equation can be put in the form
\begin{equation}
\frac{\partial\rho}{\partial t}=\hat{L}\rho\qquad\left[\textrm{Gozzi's eq. }\left(6\right)\right],\label{eq:GozziLiouvilleEquationAbstract}
\end{equation}
 where 
\begin{equation}
\hat{L}=\frac{\partial H}{\partial p}\frac{\partial}{\partial q}-\frac{\partial H}{\partial q}\frac{\partial}{\partial p},\label{eq:GozziLiouvilleOperator}
\end{equation}
 which is known as the Liouville operator (Liouville). {[}Ibid., p.
526{]}
\end{quotation}
In this paper, Gozzi did not cite any of the authors discussed in
the present work\textemdash Koopman, von Neumann, Schönberg, Loinger,
Della Riccia, Wiener, or Sudarshan.

Gozzi first brought up links between these path-integral representations
of classical mechanics and the methods of Koopman and von Neumann
in a follow-up paper in \emph{Physical Review D} in 1989, titled ``Hidden
BRS Invariance in Classical Mechanics. II'' and co-authored with
Martin Reuter and William Thacker (Gozzi, Reuter, Thacker 1989)\nocite{GozziReuterThacker:1989hbiicmi}.
Starting with this 1989 paper, Gozzi began citing Koopman's papers
and von Neumann's papers, but, again, without any mention of classical
wave functions. In the abstract, Gozzi and his co-authors wrote:
\begin{quotation}
Associated with this path integral there is an operatorial formalism
that turns out to be an extension of the well-known operatorial approach
of Liouville, Koopman, and von Neumann. {[}Ibid., p. 3363{]}
\end{quotation}
In listing the virtues of Gozzi's path-integral formulation of classical
mechanics, the authors included: 
\begin{quotation}
Second, long ago Koopman and von Neumann, influenced by the invention
of quantum mechanics, gave an \emph{operatorial} formulation of CM
{[}classical mechanics{]}. {[}Ibid., p. 3363, emphasis in the original{]}
\end{quotation}
Later on, the authors wrote:
\begin{quotation}
The crucial elements of the operatorial formalism mentioned above
are the ``classical commutation relations'' which follow from the
classical path integral. This formalism naturally embeds the standard
operator approach to CM pioneered by Liouville, Koopman and von Neumann.
{[}Ibid., p. 3364{]}
\end{quotation}
In these papers in the late 1980s and well into the 1990s, Gozzi consistently
referred to Koopman and von Neumann's method quite reasonably as ``the
operatorial approach to classical mechanics.''

Gozzi began writing papers with Danilo Mauro in the late 1990s, starting
with a preprint extending Gozzi's work on path-integral representations
of classical mechanics. This preprint appeared on the arXiv on July
9, 1999, and was eventually published in \emph{Journal of Mathematical
Physics} in 2000 (Gozzi, Mauro 2000)\nocite{GozziMauro:2000anlatsnfbonanrbfss}.
The paper began with the following introductory statement, which remained
fully in keeping with Gozzi's phrasing of Koopman and von Neumann's
methods as providing an ``operatorial'' approach to classical mechanics:
\begin{quotation}
Some time ago a \emph{path-integral} formulation of classical mechanics
(CM) appeared in the literature. This formulation was nothing else
than the path-integral counterpart of the \emph{operatorial} version
of CM provided long ago by Koopman and von Neumann. {[}Ibid., p. 1916,
emphasis in the original{]}
\end{quotation}
As with Gozzi's previous papers, this paper did not mention classical
wave functions.

Shortly thereafter, Gozzi and Mauro co-authored their next manuscript,
this time with Enrico Deotto. The preprint was titled ``Supersymmetry
in Classical Mechanics'' and showed up on the arXiv on January 18,
2001. It was published as part of a book, \emph{A Concise Encyclopaedia
of Supersymmetry}, in 2003 (Deotto, Gozzi, Mauro 2003)\nocite{DeottoGozziMauro:2003sicm}.
The article began with the following statements:
\begin{quotation}
In 1931 Koopman and von Neumann proposed an \emph{operatorial} formulation
of Classical Mechanics (CM) expanding earlier work of Liouville. Their
approach is basically the following: given a dynamical system with
a phase space $\mathcal{M}$ labelled by coordinates $\varphi^{a}=\left(q^{i},p^{i}\right)$;
$a=1,\dots,2n$; $i=1,\dots,n$, with Hamiltonian $H$ and symplectic
matrix $\omega^{ab}$, the evolution of a probability density $\rho\left(\varphi\right)$
can be given either via the Poisson brackets $\left\{ \ ,\ \right\} $
or via the Liouville operator: 
\begin{equation}
\frac{\partial\rho}{\partial t}=\left\{ H,\rho\right\} =-\hat{L}\rho;\qquad\hat{L}=\omega^{ab}\partial_{b}H\partial_{a}\qquad\left[\textrm{Deotto, Gozzi, and Mauro's eq. }\left(1\right)\right].\label{eq:DeottoGozziMauroLiouvilleEq}
\end{equation}
 The evolution via the Liouville operator is basically what is called
the operatorial approach to CM. The natural question to ask is whether
we can associate to the \emph{operatorial} formalism of CM a \emph{path
integral} one, like it is done in quantum mechanics. The answer is
yes. {[}Ibid., emphasis in the original{]}
\end{quotation}
The article cited the work of Koopman and von Neumann from the 1930s,
but, again, did not cite any of the other authors discussed in previous
sections of the present work.

\subsection{Danilo Mauro\label{subsec:Danilo-Mauro}}

On May 23, 2001, a preprint appeared on the arXiv, titled ``On Koopman\textendash von
Neumann Waves'' and authored by Danilo Mauro. The paper was later
published in \emph{International Journal of Modern Physics} in 2002
(Mauro 2002)\nocite{Mauro:2002oknw}. The paper's opening statements
were consistent with the research literature:
\begin{quotation}
In their standard formulation classical and quantum mechanics are
written in two completely different mathematical languages: for example
in classical mechanics observables are \emph{functions} of a 2n-dimensional
phase space, while in quantum mechanics they are self-adjoint \emph{operators}
acting on an {[}sic{]} Hilbert space. In the literature there are
a lot of attempts to reformulate classical and quantum mechanics in
similar forms. In this paper we shall concentrate on the work of Koopman
and von Neumann (KvN) who proposed, in 1931-32, an operatorial formulation
of classical mechanics. {[}Ibid., p. 1, emphasis in the original{]}
\end{quotation}

The statements that immediately followed, however, were historically
inaccurate, because they asserted that Koopman and von Neumann began
their approach by introducing classical wave functions: 
\begin{quotation}
The starting point of their work is the possibility of defining an
{[}sic{]} Hilbert space of \emph{complex} and \emph{square integrable}
classical ``wave'' functions $\psi\left(\varphi\right)$ such that
$\rho\left(\varphi\right)\defeq\verts{\psi\left(\varphi\right)}^{2}$
can be interpreted as a probability density of finding a particle
at the point $\varphi=\left(q,p\right)$ of the phase space. This
$\rho$ has to evolve in time according to the well-known Liouville
equation: 
\begin{equation}
i\frac{\partial}{\partial t}\rho\left(q,p\right)=\hat{\mathcal{H}}\rho\left(q,p\right)\qquad\left[\textrm{Mauro's eq. }\left(1.1\right)\right]\label{eq:MauroLiouvilleEq}
\end{equation}
 where $\hat{\mathcal{H}}$ is the Liouville operator $\hat{\mathcal{H}}=-i\partial_{p}H\partial_{q}+i\partial_{q}H\partial_{p}$
and $H$ is the Hamiltonian of the standard phase space. In order
to obtain (1.1) Koopman and von Neumann postulated the same evolution
for $\psi$: 
\begin{equation}
i\frac{\partial}{\partial t}\psi\left(q,p\right)=\hat{\mathcal{H}}\psi\left(q,p\right)\qquad\left[\textrm{Mauro's eq. }\left(1.2\right)\right]\label{eq:MauroClassicalSchrodingerEq}
\end{equation}

{[}Ibid., p. 1, emphasis in the original{]}
\end{quotation}
Notice, furthermore, that although the formula for the Liouville operator
$\hat{\mathcal{H}}$ appearing in \eqref{eq:MauroLiouvilleEq} was
in agreement with the usual definition \eqref{eq:DefLiouvillianOp},
the evolution equation \eqref{eq:MauroClassicalSchrodingerEq} for
$\psi$ was off by a crucial minus sign as compared with the equation
\eqref{eq:KoopmanFunctionTimeEvEqFromLiouvillianOp} satisfied by
Koopman's phase-space functions.

Later, after reviewing the equations for time evolution in textbook
quantum mechanics, the paper continued as follows:
\begin{quotation}
The situation is completely different in the Hilbert space of classical
mechanics. In fact, following Koopman and von Neumann, we postulate
that the wave functions $\psi\left(\varphi,t\right)=\psi\left(q,p,t\right)$
evolve in time with the Liouvillian operator: 
\begin{equation}
\hat{\mathcal{H}}=-i\partial_{p_{i}}H\partial_{q_{i}}+i\partial_{q_{i}}H\partial_{p_{i}}\qquad\left[\textrm{Mauro's eq. }\left(2.5\right)\right]\label{eq:MauroDefLiouvillianOperator}
\end{equation}
 according to the following equation: 
\begin{equation}
i\frac{\partial}{\partial t}\psi=\hat{\mathcal{H}}\psi\quad\Rightarrow\quad\frac{\partial}{\partial t}\psi=\left(-\partial_{p_{i}}H\partial_{q_{i}}+\partial_{q_{i}}H\partial_{p_{i}}\right)\psi\qquad\left[\textrm{Mauro's eq. }\left(2.6\right)\right]\label{eq:MauroLiouvilleEqForWaveFunction}
\end{equation}
 We can think of (2.6) as the analogue of the quantum Schrödinger
equation, i.e. as the fundamental equation governing the evolution
of the vectors in the Hilbert space of classical mechanics. These
vectors are the complex wave functions on the phase space obeying
the normalizability condition $\int dqdp\,\psi^{\ast}\left(q,p\right)\psi\left(q,p\right)=1$.
{[}Ibid., p. 3{]}
\end{quotation}
As the present work has established, Koopman and von Neumann did not
introduce classical wave functions, nor did they postulate that any
such classical wave functions evolved according to the Liouville equation.
These ideas were, in fact, originally due to Schönberg, and then developed,
in some cases independently, by other researchers, including Loinger,
Della Riccia, Wiener, and Sudarshan. As such, this method of classical
wave functions should not properly be called ``Koopman\textendash von
Neumann classical mechanics.''

This incorrect history showed up again in a preprint authored by Mauro
and Gozzi that appeared on the arXiv on the same day\textemdash May
23, 2001. The preprint was titled ``Minimal Coupling in Koopman\textendash von
Neumann Theory,'' and was published in \emph{Annals of Physics} in
2002 (Gozzi, Mauro 2002)\nocite{GozziMauro:2002mcikvnt}. The opening
remarks began with: 
\begin{quotation}
In 1931, Koopman and von Neumann (KvN) \emph{postulated} the same
evolution equation for complex distributions $\psi\left(q,p\right)$
making up an $L^{2}$ Hilbert space: 
\begin{equation}
\partial_{t}\psi\left(q,p\right)=-\hat{L}\psi\left(q,p\right)\qquad\left[\textrm{Gozzi and Mauro's eq. }\left(1.2\right)\right].\label{eq:GozziMauroLiouvilleEqWaveFunction}
\end{equation}
 If we postulate {[}this equation{]} for $\psi\left(q,p\right)$,
then it is easy to prove that functions $\rho$ of the form 
\begin{equation}
\rho=\verts{\psi}^{2}\qquad\left[\textrm{Gozzi and Mauro's eq. }\left(1.3\right)\right]\label{eq:GozziMauroProbabiltyDensityAsModSqWaveFunction}
\end{equation}
 evolve with the same equation as $\psi$. This is so because the
operator $\hat{L}$ contains only firs {[}sic{]} order derivatives.
This is not what happens in quantum mechanics (QM) where the evolution
of the $\psi\left(q\right)$ is via the Schrödinger operator $\hat{H}$
while that of the associated $\rho=\verts{\psi}^{2}$ is via a totally
different operator. The reason is that the Schrödinger operator $\hat{H}$,
differently than {[}sic{]} the Liouville operator $\hat{L}$, contains
second order derivatives. {[}Ibid., pp. 152\textendash 153, emphasis
in the original{]}
\end{quotation}
An equation resembling \eqref{eq:GozziMauroLiouvilleEqWaveFunction}
does appear in Koopman's 1931 paper, though it is intended for describing
the time evolution of observables in a classical phase space. The
next statement in the 2002 paper is also historically incorrect: 
\begin{quotation}
By postulating the relations (1.3) and (1.2) for the $\psi$, KvN
managed to build an operatorial formulation for classical mechanics
(CM) equipped with a Hilbert space structure and producing the same
results as the Liouville formulation. {[}Ibid.{]}
\end{quotation}
Koopman and von Neumann did not postulate either of Gozzi and Mauro's
equations (1.2) or (1.3). As with Mauro's previous paper, the 2002
paper cites Koopman and von Neumann, but does not cite Schönberg or
any of the other developers of the method of classical wave functions.

Soon after, Deotto, Gozzi, and Mauro collaborated on another paper,
which appeared on the arXiv on August 7, 2002. The paper, titled ``Hilbert
Space Structures in Classical Mechanics. I,'' was published in \emph{Journal
of Mathematical Physics} in 2003 (Deotto, Gozzi, Mauro 2003)\nocite{DeottoGozziMauro:2003hssicmi}.
The paper began with the following statements:
\begin{quotation}
In the 1930s Koopman and von Neumann (KvN) gave an operatorial formulation
of \emph{classical mechanics} (CM). They first introduced square-integrable
functions $\psi\left(\varphi^{a}\right)$ on the phase space $\mathcal{M}$
of a classical system with Hamiltonian $H\left(\varphi\right)$ (with
$\varphi^{a}$ we indicate the 2$n$ phase-space coordinates of the
system $\varphi^{a}=q^{1}\cdots q^{n},p^{1}\cdots p^{n}$). According
to KvN the Liouville phase-space distributions are obtained from $\psi\left(\varphi\right)$
as 
\begin{equation}
\rho\left(\varphi\right)=\verts{\psi\left(\varphi\right)}^{2}\qquad\left[\textrm{Deotto, Gozzi, and Mauro's eq. }\left(1.1\right)\right].\label{eq:DeottoGozziMauroProbabilityDensityFromModSq}
\end{equation}
 The introduction of the $\psi\left(\varphi\right)$ is an acceptable
assumption considering that $\rho\left(\varphi\right)$, having the
meaning of a probability density, is always positive semidefinite
$\rho\left(\varphi\right)\geqslant0$, and so one can always take
its ``square root'' and obtain $\psi\left(\varphi\right)$. Moreover,
as $\psi\left(\varphi\right)$ is square integrable, i.e., $\psi\left(\varphi\right)\in L^{2}$,
it turns out that $\rho\left(\varphi\right)$ is integrable as it
should be 
\begin{equation}
\int\mathrm{d}^{2n}\varphi\,\psi^{\ast}\left(\varphi\right)\psi\left(\varphi\right)=\int\mathrm{d}^{2n}\varphi\,\rho\left(\varphi\right)<\infty\qquad\left[\textrm{Deotto, Gozzi, and Mauro's eq. }\left(1.2\right)\right].\label{eq:DeottoGozziMauroNormalization}
\end{equation}
 KvN \emph{postulated} the following evolution for $\psi\left(\varphi\right)$:
\begin{equation}
i\frac{\partial\psi\left(\varphi,t\right)}{\partial t}=\hat{L}\psi\left(\varphi,t\right)\qquad\left[\textrm{Deotto, Gozzi, and Mauro's eq. }\left(1.3\right)\right]\label{eq:DeottoGozziMauroLiouvillEqWaveFunction}
\end{equation}
 where $\hat{L}$, defined as 
\begin{equation}
\hat{L}=i\frac{\partial H}{\partial q^{i}}\frac{\partial}{\partial p^{i}}-i\frac{\partial H}{\partial p^{i}}\frac{\partial}{\partial q^{i}}\qquad\left[\textrm{Deotto, Gozzi, and Mauro's eq. }\left(1.4\right)\right]\label{eq:DeottoGozziMauroLiouvillianOp}
\end{equation}
 is the Liouville operator. This equation of motion for $\psi\left(\varphi\right)$
and (1.1) lead to the same evolution for $\rho\left(\varphi\right)$,
\begin{equation}
i\frac{\partial\rho\left(\varphi,t\right)}{\partial t}=\hat{L}\rho\left(\varphi,t\right)\qquad\left[\textrm{Deotto, Gozzi, and Mauro's eq. }\left(1.5\right)\right].\label{eq:DeottoGozziMauroLiouvillEqProbabilityDensity}
\end{equation}
 This is the well-known Liouville equation satisfied by the classical
probability densities. Note that $\rho\left(\varphi\right)$ obeys
the same equation as $\psi\left(\varphi\right)$ because $\hat{L}$
is first order in the derivatives. The same does not happen in quantum
mechanics where the analog of (1.3) is the Schrödinger equation whose
evolution operator is second order in the derivatives. We will not
spend more time here in explaining the interplay between the quantum
mechanical wave functions $\psi\left(q\right)$ and these \textquoteleft\textquoteleft KvN
waves\textquoteright\textquoteright{} $\psi\left(\varphi\right)$.
The interested reader can consult Ref. 2 {[}which refers to the two
2002 papers cited above{]} where many details have been worked out.
{[}Ibid., pp. 5902\textendash 5903, emphasis in the original{]}
\end{quotation}
Again, the paper attributed the introduction of classical wave functions,
related to classical probability distributions by the modulus-squaring
operation, and the introduction of a time-evolution equation for classical
wave functions, to Koopman and von Neumann.

Mauro's PhD thesis appeared on the arXiv on January 30, 2003, and
was titled ``Topics in Koopman\textendash von Neumann Theory'' (Mauro
2003)\nocite{Mauro:2003tikvnt}. The introduction, in describing attempts
to connect the formalism of classical mechanics with the formalism
of quantum mechanics, contained this text: 
\begin{quotation}
Another, even older, direction is to reformulate CM in an operatorial
language by using a Hilbert space of square integrable functions on
the phase space and by replacing the Poisson brackets with some suitable
classical commutators. This is what has been done in the 30\textquoteright s
by Koopman and von Neumann (KvN). {[}Ibid., p. 1{]}
\end{quotation}
These statements were consistent with the historical record. However,
a few sentences later, one finds: 
\begin{quotation}
The starting point of KvN is the introduction of a Hilbert space of
square integrable and complex functions $\psi\left(q,p\right)$ whose
modulus square are just the usual probability densities in phase space
$\rho\left(q,p\right)=\verts{\psi\left(q,p\right)}^{2}$. {[}Ibid.,
p. 1{]}
\end{quotation}
Shortly thereafter, the paper stated: 
\begin{quotation}
In fact KvN postulated that the evolution of the $\psi\left(\varphi\right)$
must be given by the Liouvillian which, containing only first order
derivatives, evolves also the probability densities $\rho\left(\varphi\right)$,
differently than {[}sic{]} what happens in QM. {[}Ibid., pp. 1\textendash 2{]}
\end{quotation}
As the present work has shown, these last two statements were not
in keeping with the historical record. Similar statements showed up
elsewhere in the thesis, in which $\psi\left(\varphi\right)$ was
called the ``KvN wave'' and its Liouville equation was called the
``KvN equation.'' Although the thesis did not cite most of the researchers
discussed in the present work, the thesis did cite Sherry and Sudarshan's
1978 paper (Sherry, Sudarshan 1978)\nocite{SherrySudarshan:1978ibcaqsanatqmi},
albeit without crediting the use of classical wave functions to Sudarshan.

The introduction to Mauro's 2003 paper ``A New Quantization Map''
(Mauro 2003)\nocite{Mauro:2003anqm} contained this text: 
\begin{quotation}
KvN formulated classical mechanics in a Hilbert space made up of complex
square integrable functions over the phase space variables $\psi\left(q,p,t\right)$.
In particular they postulated, as equation of evolution for $\psi\left(q,p,t\right)$,
the Liouville equation itself 
\begin{equation}
i\frac{\partial}{\partial t}\psi\left(q,p,t\right)=\hat{L}\psi\left(q,p,t\right)\qquad\left[\textrm{Mauro's eq. }(3)\right].\label{eq:MauroLiouvilleEqForWaveFunctioNewQuantization}
\end{equation}
 Starting from (3) it is easy to prove that, since the Liouvillian
$\hat{L}$ contains only first order derivatives, the Liouville equation
(1) for the probability densities $\rho\left(q,p,t\right)$ can be
derived via the postulate $\rho\left(q,p,t\right)=\verts{\psi\left(q,p,t\right)}^{2}$.
Finally KvN imposed on the states of their Hilbert space the following
scalar product: 
\begin{equation}
\langle\psi\vert\tau\rangle=\int dq\,dp\,\psi^{\ast}\left(q,p\right)\tau\left(q,p\right)\qquad\left[\textrm{Mauro's eq. }(4)\right].\label{eq:MauroInnerProductStates}
\end{equation}
 With this choice the Liouvillian $\hat{L}$ is a Hermitian operator.
Therefore $\langle\psi\vert\psi\rangle=\int dq\,dp\,\verts{\psi\left(q,p\right)}^{2}$
is a conserved quantity and $\verts{\psi\left(q,p\right)}^{2}$ can
be consistently interpreted as the probability density of finding
a particle in a point of the phase space. {[}Ibid., p. 28{]}
\end{quotation}
Despite the incorrect claim that Koopman and von Neumann postulated
the specific equation \eqref{eq:MauroLiouvilleEqForWaveFunctioNewQuantization},
notice the careful language in these introductory statements, which
otherwise said only that Koopman and von Neumann identified $\psi\left(q,p,t\right)$
as ``complex square integral functions over the phase space variables''
with a specific scalar product, and did not claim that Koopman or
von Neumann themselves interpreted $\psi\left(q,p,t\right)$ as a
classical wave function. In the quoted text above, the paper switched
to the passive voice in assigning this interpretation to $\psi\left(q,p,t\right)$,
both in the sentence immediately following \eqref{eq:MauroLiouvilleEqForWaveFunctioNewQuantization},
and in the last sentence of the quoted text. The scalar product appearing
in \eqref{eq:MauroInnerProductStates}, however, was missing the phase-space
probability density $\rho\left(q,p\right)$ that appeared in the inner
product \eqref{eq:DefKoopmanInnerProduct} as defined in Koopman's
1931 paper, and that also appeared in the integral measures used both
in Koopman's 1931 paper and in von Neumann's two 1932 papers.

Similar statements showed up in a 2004 paper co-authored by Gozzi
and Mauro, titled ``On Koopman\textendash von Neumann Waves II''
(Gozzi, Mauro 2004). In the abstract, the authors wrote:
\begin{quotation}
In particular we show that the introduction of the KvN Hilbert space
of complex and square integrable \textquotedblleft wave functions\textquotedblright{}
requires an enlargement of the set of the observables of ordinary
classical mechanics. {[}Ibid.{]}
\end{quotation}
The introduction included:
\begin{quotation}
In a previous paper we stressed that KvN did not use the space of
the $\rho$ but introduced instead a Hilbert space made up of complex
\emph{square integrable} functions $\psi\left(q,p\right)\in L^{2}$
over phase space. These $\psi$ are \textquotedblleft the KvN waves\textquotedblright{}
we indicated in the title. Next they \emph{postulated} for every $\psi\left(q,p,t\right)$
an equation of evolution which is the Liouville equation itself: 
\begin{equation}
i\frac{\partial}{\partial t}\psi\left(q,p,t\right)=\hat{\mathcal{H}}\psi\left(q,p,t\right)\qquad\left[\textrm{Gozzi and Mauro's eq. }\left(1.3\right)\right].\label{eq:GozziMauroKvNWavesIILiouvilleEqForWaveFunction}
\end{equation}
 Because the Liouvillian $\hat{\mathcal{H}}$ contains only first
order derivatives, it is easy to prove that the Liouville equation
(1.1) for the probability densities $\rho\left(q,p,t\right)$ can
be derived from (1.3) by postulating that $\rho\left(q,p,t\right)=\verts{\psi\left(q,p,t\right)}^{2}$.
{[}Ibid., emphasis in the original{]}
\end{quotation}
Although the paper did not attribute the modulus-squaring relationship
to Koopman and von Neumann, it did refer to $\psi$ as a ``KvN wave''
and said that Koopman and von Neumann postulated its time-evolution
equation.

\section{Conclusion\label{sec:Conclusion}}

The present paper was intended to clarify the history behind the formulation
of classical mechanics developed by Bernard Koopman and John von Neumann,
and to show that the method of ``classical'' wave functions was
due not to Koopman and von Neumann, but perhaps first due to Mario
Schönberg, with later contributions from Alfred Loinger, Giacomo Della
Riccia, Norbert Wiener, and E.C. George Sudarshan. Of course, even
this historical assessment may be incorrect, and other researchers
may have come up with the idea before Schönberg.

The method of classical wave functions has proved to be important
and highly useful, and this historical misattribution should not be
taken to suggest otherwise. However, getting the history right matters,
especially when those responsible for an influential idea would otherwise
not be given the credit that they deserve.

\section*{Acknowledgments}

The author would especially like to thank Ennio Gozzi, John Norton,
and Miklos Redei for helpful and informative communications.

\bibliographystyle{1_home_jacob_Documents_Work_My_Papers_2025-Koop____Neumann_Misattributed_custom-abbrvalphaurl}
\bibliography{0_home_jacob_Documents_Work_My_Papers_Bibliography_Global-Bibliography}

\end{document}